\documentclass[12pt,preprint]{aastex}

\newcommand{\hst}{\textit{HST\ }}

\begin{document}
\title{First Constraints on Rings in the Pluto System}

\author{A.~J.~Steffl and S.~A.~Stern}
\email{steffl@boulder.swri.edu}

\affil{Southwest Research Institute, Space Science and Engineering
  Division}
\affil{1050 Walnut Street, Suite 400, Boulder, CO 80302}

\shorttitle{Constraints on Pluto's Rings}
\shortauthors{Steffl and Stern}
\slugcomment{Submitted to the Astronomical Journal}

\begin{abstract}
  Simple theoretical calculations have suggested that small body
  impacts onto Pluto's newly discovered small satellites, Nix and
  Hydra, are capable of generating time-variable rings or dust sheets
  in the Pluto system. Using HST/ACS data obtained on 2006 February 15
  and 2006 March 2, we find no observational evidence for such a ring
  system and present the first constraints on the present-day I/F and
  optical depth of a putative ring system. At the 1500-km radial
  resolution of our search, we place a 3$\sigma$ upper limit on the
  azimuthally-averaged normal I/F of ring particles of 5.1x10$^{-7}$
  at a distance of 42,000~km from the Pluto-Charon barycenter, the
  minimum distance for a dynamically stable ring \citep{Sternetal94,
    Nagyetal06}; 4.4x10$^{-7}$ at the orbit of Nix; and 2.5x10$^{-7}$
  at the orbit of Hydra. For an assumed ring particle albedo of 0.04
  (0.38), these I/F limits translate into 3$\sigma$ upper limits on
  the normal optical depth of macroscopic ring particles of
  1.3x10$^{-5}$ (1.4x10$^{-6}$), 1.1x10$^{-5}$ (1.2x10$^{-6}$),
  6.4x10$^{-6}$ (6.7x10$^{-7}$), respectively. Were the {\it New
    Horizons} spacecraft to fly through a ring system with optical
  depth of 1.3x10$^{-5}$, it would collide with a significant number
  of potentially damaging ring particles. We therefore recommend that
  unless tighter constraints can be obtained, {\it New Horizons} cross
  the putative ring plane within 42,000~km of the Pluto-Charon
  barycenter, where rings are dynamically unstable. We derive a crude
  estimate of the lifetime of putative ring paritcles of 900 years.

\end{abstract}


\section{Introduction}

The discovery of Pluto's two small ($\sim$100~km diameter) satellites,
Nix and Hydra, in 2005 \citep{Weaveretal06} raised the possibility
that Pluto may possess a time-variable ring system
\citep{Sternetal06}. Previously, \cite{Durda:stern00} demonstrated
that collisions between small Kuiper belt debris and larger objects,
such as Pluto's satellites, are a common occurrence. The
characteristic ejecta velocity resulting from these collisions will be
of order 1--10\% of the impactor velocity, or 10--100~m s$^{-1}$.
Since the escape velocities of Nix and Hydra are between 30 and 90~m
s$^{-1}$, depending on their precise size and density, ejecta from
such collisions can escape from Pluto's small satellites but will
generally remain gravitationally bound to Pluto, thus forming rings.
This is in contrast to the situation at Charon, where most such ejecta
will fall back onto the surface, due to Charon's $\sim$500~m s$^{-1}$
escape velocity. Such rings will be highly time-variable, as
collisions on Nix and Hydra compete with loss processes.
\cite{Sternetal06} calculated a characteristic ring optical depth of
$\tau$=5x10$^{-6}$ for ring material between Nix and Hydra. This is
considerably more tenuous than the main rings of Saturn, Uranus, and
Neptune, but comparable to the optical depth of Jupiter's rings
\citep{Burnsetal84} and the faint dust rings of Uranus
\citep{Showalter:lissauer06}.

No prior observational constraints on the reflectance or optical depth
of rings in the Pluto system have been published. We therefore
undertook a search for such material using existing datasets acquired
by the \hst Advanced Camera for Surveys (ACS) in early 2006. In
addition to the discovery potential of this search, detection of or
constraints on Plutonian rings can yield useful information about both
the potential hazards of ring material to the NASA {\it New Horizons}
Pluto mission, now en route, and the small body population orbiting in
the 30--50 AU region of the Kuiper belt traversed by Pluto.

In what follows we describe the dataset we used to constrain the
amount of ring material in the Pluto system, the analysis techniques
we employed, and the results we obtained. We then go on to interpret
our results, which are upper limits, as they apply to ring system
hazards for {\it New Horizons} and the lifetime of ring particles.

\section{Observations}

Immediately following the discovery of Nix and Hydra, our team
obtained Director's Discretionary time to use \hst to confirm the
existence of the new satellites. However, to conserve the lifetime of
the remaining functional \hst gyros, \hst began operations in 2-gyro
mode before the confirmation observations could be executed. In this
mode, most targets can only be observed when they are on the trailing
side of the Sun as it moves along the ecliptic, thus precluding
observations of the Pluto system until February 2006. When Pluto again
became observable, it was targeted with two \hst visits of one orbit
each using the ACS High Resolution Channel (HRC) through Guest
Observer/Director's Discretionary (GO/DD) program 10774. Selected
observational parameters from these visits are presented in
Table~\ref{obs_table}.  \placetable{obs_table}

The first visit occurred on 2006 Feb 15.7 UT. During this visit, four
long exposures (475-second integrations) with the F606W filter (broad
V) were obtained using a non-integer dither box pattern to minimize
the effect of bad detector pixels and improve the spatial sampling. In
these long exposures, both Pluto and Charon are saturated with a small
amount of bleeding along the CCD columns. To aid the registration of
the long exposures, short, unsaturated exposures (1-second
integrations) were also obtained at each point in the dither pattern.
Both Nix and Hydra were clearly detected at high signal to noise
\citep{Mutchleretal06IAUC}, and no additional satellites were detected
between the orbits of Nix and Hydra down to a 90\%-confidence limiting
magnitude of V=25.7 \citep{Steffletal06b}.

With the successful confirmation of both satellites from analysis of
archival ACS data \citep{Buieetal06} and the 2006 February 15 visit,
the observational setup of the second visit, which occurred on 2006
March 2, was modified to also obtain the first \bv colors of Nix and
Hydra \citep{Sternetal06barxiv}. A three-point, non-integer dither
pattern was employed with a 145~s exposure using the F606W filter and
a 475~s exposure using the F435W (Johnson B) obtained at each point of
the dither pattern. A 3~s exposure using the F435W filter and a 1~s
exposure using the F606W filter were also obtained at points 1 and 3
of the dither pattern, respectively, to aid in image registration.
Since the point spread function (PSF) of the ACS varies significantly
between filters, data taken during the 2006 March 2 visit with the
F606W filter were analyzed separately from data taken with the F435W
filter.

\begin{figure}
  \plotone{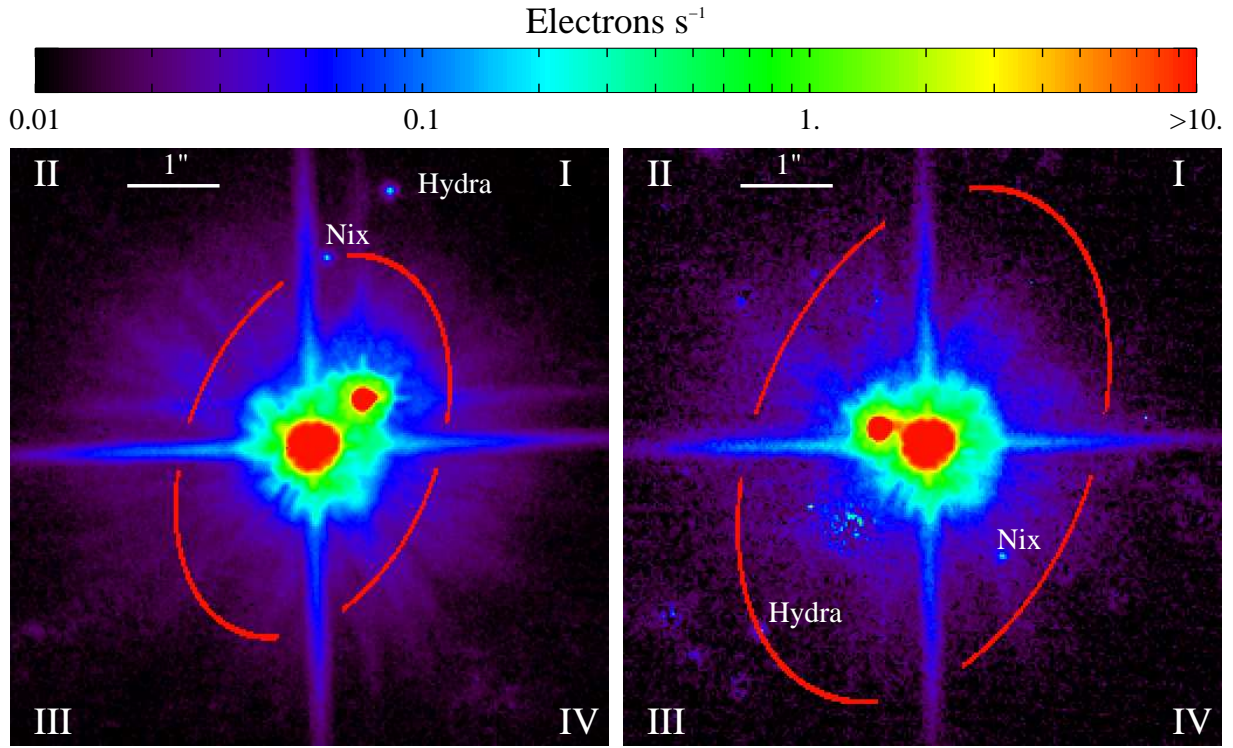} 
  \caption[]{Drizzled HRC F606W images of the Pluto system. The image 
    on the left is from the 2006 February 15 visit, while the image on
    the right is from the 2006 March 2 visit. Pluto, Charon, Nix, and
    Hydra can be clearly seen. All other features are due to the
    extended PSF halos around Pluto and Charon or artifacts introduced
    by the incomplete removal of background star trails or cosmic ray
    events.  The diffraction spikes from Pluto have been used to
    divide each image into four quadrants. 1500~km width annuli at the
    orbital distances of Nix (left) and Hydra (right) are shown in
    red. Four 10$^{\circ}$ segments (half width), centered on each
    diffraction spike, are excluded from analysis. A one arcsecond
    scale bar is shown in the upper left of each image.
 \label{plutoim}} 
\end{figure}

For each visit, the exposures made using a given filter were
``drizzled'' together using the PyRAF {\it Multidrizzle} procedure
\citep{Koekemoeretal02}. Via this procedure, the individual images are
corrected for the geometric distortion of the ACS instrument, rotated
so that north is up and east to the left, sky background subtracted,
co-registered relative to Pluto using the non-saturated 1-s exposures,
and combined using a median filter. Detector pixels that have
anomalously low sensitivity, high dark counts, or are saturated were
flagged and excluded from further analysis. The median combination
removes artifacts, such as cosmic ray events or star trails, that do
not appear in same position relative to Pluto in at least two of the
images. The resulting images from the February and March visits using
the F606W filter are shown in Figure~\ref{plutoim}.
\placefigure{plutoim}

Pluto is highly saturated in the long integrations obtained during
both visits. Charon is also saturated in the long integrations
obtained during the February visit. As a result, the two-dimensional
pattern of flux from Pluto and Charon dominates over the sky
background, even several arcseconds away from Pluto. Complex structure
can be seen in this pattern, as a result of the higher-order behavior
in the wings of the ACS HRC point spread function (PSF).

\section{Results}

To reduce the strong background gradients in the processed images due
to Pluto and Charon, we employ the high-pass filter method of
\cite{Showalter:lissauer06}. At each pixel in the image, a
first-order, two-dimensional polynomial (i.e. a tilted plane) is fit
to the pixels lying within a surrounding circular region. A small
circular aperture at the center of this region is excluded from the
fit so as not to reduce the intensity of point features. A radius of
12 pixels was chosen for the outer limit of the fit; a radius of 5
pixels was chosen for the inner aperture, to enclose 80\% of the flux
from a point source \citep{Siriannietal05}.  The value of the fit is
then subtracted from the central pixel and the process repeated for
all pixels in the drizzled images. The resulting filtered images still
contain the high-frequency components of the 2-D pattern of flux from
Pluto and Charon, but the low-frequency components are much reduced.
Visual inspection of the filtered images reveals no immediate evidence
for a faint ring system.  As noted by \cite{Showalter:lissauer06},
this high-pass filter method is not photometrically accurate (tending
to narrow bright features), so an alternative method must be used to
place quantitative limits on the I/F of a potential ring system.

Flux from Pluto and Charon is the dominant source of counts in the
region of the images where a ring system might exist.  To zeroth
order, the PSFs of the ACS HRC are azimuthally symmetric.  However,
the long integration times used in these observations allow the
non-symmetric components of the PSF, i.e., the wings, to rise
significantly above the level of the sky background.  The situation is
also complicated by the fact that neither Pluto nor Charon are true
point sources, having apparent angular diameters of 0\farcs104 and and
0\farcs052, respectively, during the February visit (for comparison,
the pixel size of the ACS HRC is approximately 0\farcs025). Pluto is
also known to exhibit significant albedo variations across its surface
\citep{Sternetal97, Youngetal01}. As a result, the observed pattern of
flux from Pluto and Charon can not be accurately removed by the simple
subtraction of an azimuthally-averaged image from the data. In
principle, more sophisticated techniques, e.g. PSF subtraction, can be
used to remove this pattern; however, outside of the PSF core, the
observed non-symmetric, high-frequency variations in flux from Pluto
and Charon are poorly described by model ACS PSFs, such as those
generated by the Tiny Tim program \citep{Krist:hook04}.  Since neither
the 2-D pattern of flux from Pluto and Charon nor the spatial extent
of a putative ring system are known {\it a priori}, attempts to
characterize and remove Pluto and Charon's flux using fits to the data
run a significant risk of inadvertently removing some or all of the
signal from a potential ring system. Thus, given the difficulty of
removing the background flux from Pluto and Charon without introducing
artifacts to the data, we have adopted a simple approach: assuming
that all of the flux contained in the image is due to light
backscattered from a ring system. This is clearly an
oversimplification, but it yields a firm upper limit on the
reflectance of a putative ring system.

Since the orbits of Pluto's three known moons are all essentially
co-planar \citep{Weaveretal06,Buieetal06}, any ring system would be
expected to share this plane. We therefore divided the plane of
Charon's orbit into 100 circular annuli centered on the Pluto-Charon
barycenter. We chose a width of 1500~km for the annuli, based on the
projected size of three of the 0\farcs025 HRC pixels at Pluto. Our
results, however, are only weakly dependent on the choice of annulus
width. Areas of the image containing diffraction spikes from either
Pluto or Charon were masked out and excluded from further analysis, as
were circular regions surrounding Pluto, Charon, Nix, and Hydra.
Pluto's diffraction spikes provide a natural means of dividing the
image into quadrants. Quadrant boundaries were used to divide each
annulus into four segments. The total count rate contained in and
solid angle subtended by each annulus segment within a given quadrant
was then calculated.

The standard measurement of reflectance for planetary rings is the
dimensionless ratio I/F, where I is the observed intensity and $\pi$F
is the incident (solar) flux. For an optically thin ring, I/F is
proportional to 1/sin(B), where B is the ring opening angle, defined
as the angle between the line of sight and the ring plane. We
therefore multiply I/F by sin(B) to remove this geomterical factor,
yielding the ``normal I/F'', i.e., that seen from an observer looking
perpendicular to the ring plane at the same phase angle. The normal
I/F is given by the following equation:

\begin{equation}
  \label{I_over_F_eqn}
  \left(\frac{I}{F}\right)_{\perp}= \frac{C r^2 \sin(B)}{\pi \Omega} 
  10^{(m_{\sun}+\Delta m (\alpha)-z)/2.5} 
\end{equation}

\noindent where $C$ is the total count rate observed in the 
segment(s) of an annulus lying within a particular quadrant, in
electrons s$^{-1}$; $r$ is the distance from the Sun to Pluto in AU;
$\Omega$ is the solid angle subtended by the annulus segment(s); $B$
is the ring opening angle, defined as the angle between the line of
sight and the ring plane; $m_{\sun}=-26.70$ is the apparent magnitude
of the Sun at a distance of 1~AU using the HRC F606W filter bandpass
\citep{Colinaetal96}; $\Delta m (\alpha)$ is the difference in
magnitude between opposition and phase angle $\alpha$; and $z=26.13$
is the magnitude system zero point \citep{Siriannietal05}. Since the
phase function of the putative ring system is unknown, we have adopted
the phase function used by \cite{Karkoschka01a} to model the rings of
Uranus:

\begin{equation}
  \label{phase_func_eqn}
  \Delta m (\alpha)=\beta \alpha+0.5\alpha/(\alpha_0+\alpha)
\end{equation}

\noindent The first term of this equation is the standard linear 
dependence with a phase coefficient of $\beta=0.03$ mag deg$^{-1}$.
The second term represents the opposition surge with a width at half
maximum of $\alpha_0=8$ deg$^{-1}$. For the phase angle of
$\alpha$=1.69$^{\circ}$ during the February visit, these two terms
result in a correction of 0.14 mag.

If Pluto's rings are more tightly confined radially than the
$\sim$1500~km resolution of our search, Eq.~\ref{I_over_F_eqn} will
underestimate the true I/F. However, given the apparent lack of any
additional small satellites in the Pluto system \citep{Steffletal06b},
it is difficult to see how a ring system could remain so tightly
confined radially.

\begin{figure}\epsscale{.8} 
  \plotone{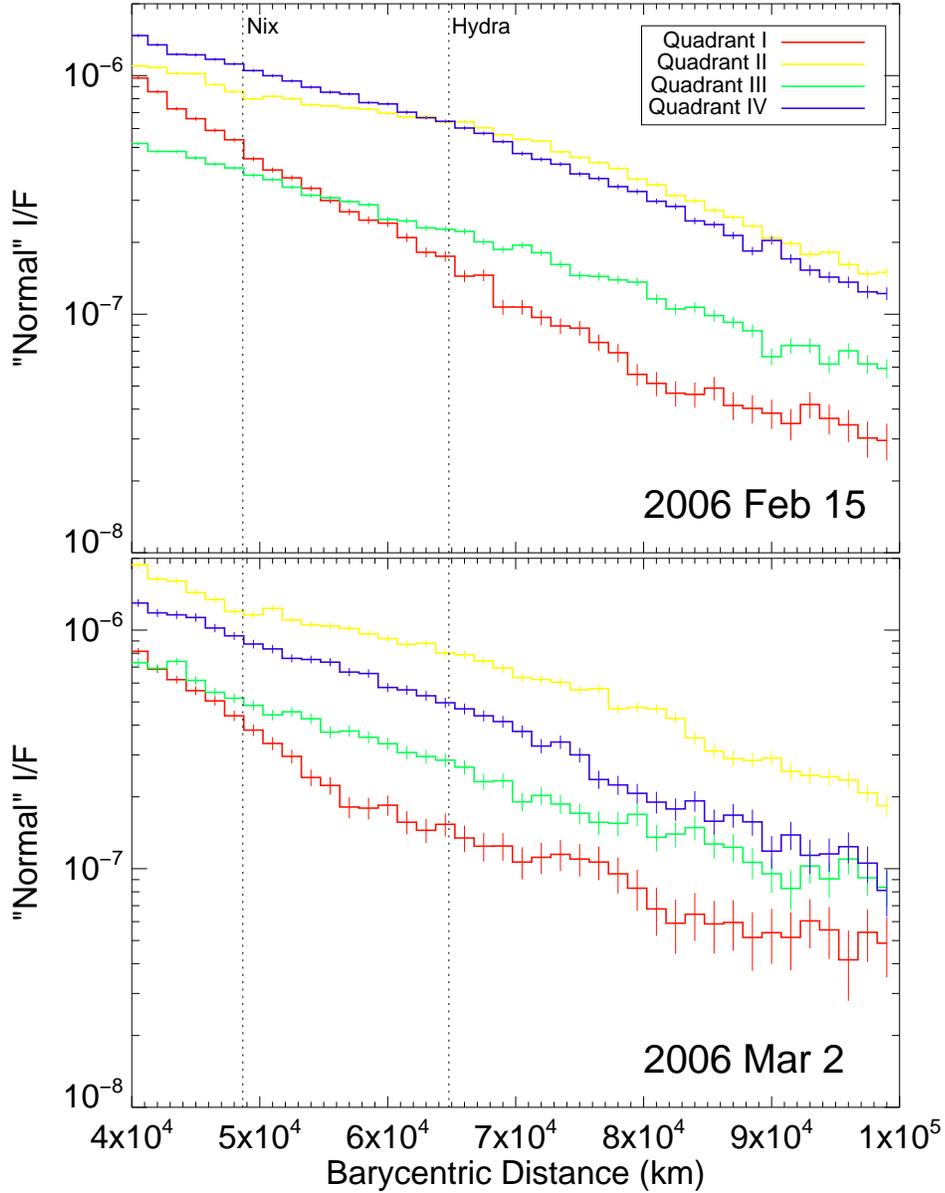}
\epsscale{1}
  \caption[]{``Normal'' I/F as a function of barycentric distance, 
    assuming all observed flux is backscattered from a ring system.
    1$\sigma$ error bars are shown at each point. For a given
    barycentric distance, annulus segments in quadrants I and III
    appear at a greater angular distance from Pluto, owing to the
    projection of the putative ring plane on the plane of the sky.
    Since the observed flux in a given annulus segment is dominated by
    flux from Pluto, these quadrants provide the tightest constraints
    on the I/F of a putative ring system. Normal I/F values shown for
    the 2006 March 2 visit are derived exclusively from data using the
    F606W filter, as the normal I/F values derived from the F435W data
    are significantly higher. The orbital semi-major axes of Nix and
    Hydra are shown by vertical dotted lines at 48,675 and 64780 km,
    respectively.
 \label{ioverf_vs_radius}} 
\end{figure}

Profiles of the normal I/F versus distance from the Pluto-Charon
barycenter, as derived from the two \hst visits, are shown in
Figure~\ref{ioverf_vs_radius}. In all quadrants, the derived I/F value
falls off sharply with increasing distance from Pluto, evidence that
light from Pluto dominates the total flux contained in the annulus
segments. There is no statistically significant evidence for a ring
system in any of the I/F profiles. Profiles from the 2006 March 2
visit using the F435W filter have a significantly higher I/F value at
a given barycentric distance (owing to the broader PSF with this
filter), and therefore were not used in further analysis.

The inclination of the line of sight to the orbital plane of Pluto's
moons causes the circular annuli to appear elliptical when projected
onto the plane of the sky. As a result, annulus segments appear closer
to Pluto in quadrants II and IV than in quadrants I and III. Thus, I/F
upper limits derived from these quadrants are significantly higher, as
is readily seen in Figure~\ref{plutoim}.  During the 2006 February 15
visit, quadrant III provides the tightest constraint on the normal I/F
of a putative ring system between the orbits of Nix and Hydra, while
during the 2006 March 2 visit, quadrant I provides the tightest
constraint. However, the statistical error associated with the I/F
profile from quadrant III of the February visit is significantly lower
than that from quadrant I of the F606W data from the March visit,
owing to the factor of 4 greater integration time and better rejection
of star trails and cosmic rays events during the February visit.

Dynamical interactions with Charon render orbits about the
Pluto-Charon barycenter with semi-major axes less than 42,000~km (2.15
times the orbital semi-major axis of Charon) unstable
\citep{Sternetal94, Nagyetal06}. The 3$\sigma$ upper limits on the
normal I/F of a putative ring are given at this inner boundary and the
orbits of Nix and Hydra in Table~\ref{constraints_table}.
\placefigure{ioverf_vs_radius} \placetable{constraints_table}

\begin{figure} 
  \plotone{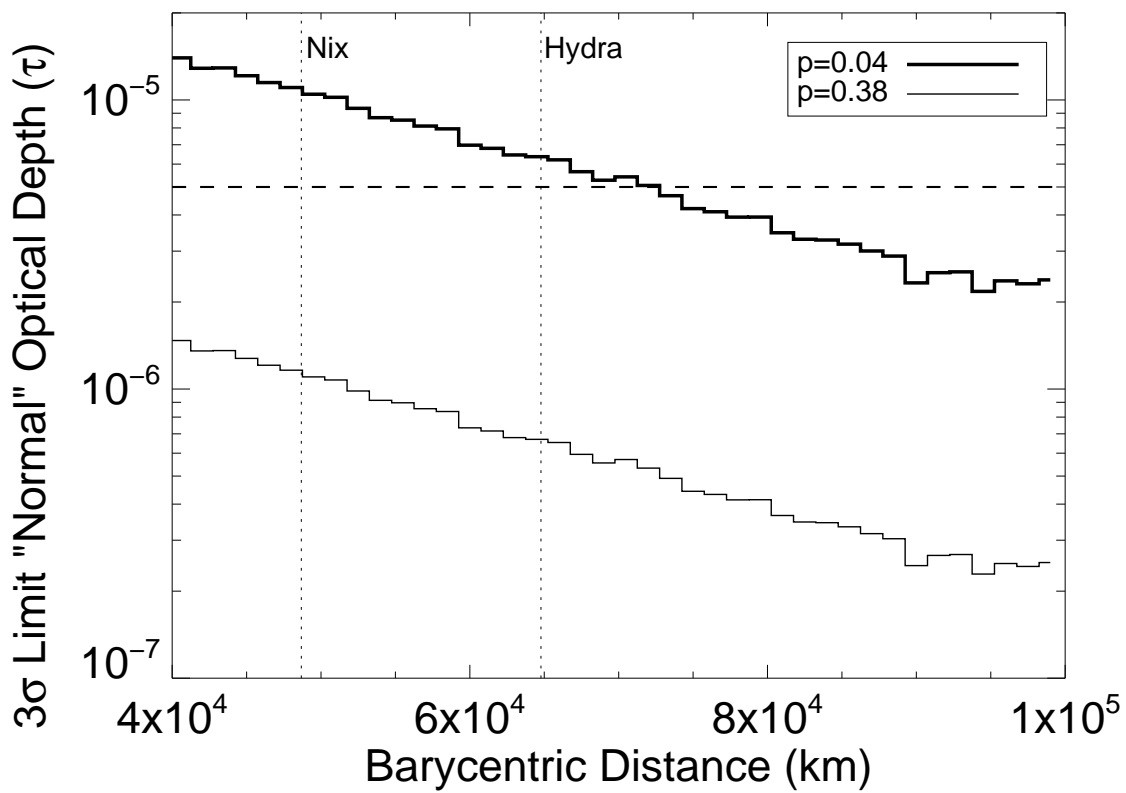}
  \caption[]{3$\sigma$ upper limit on ``normal'' optical depth for a 
    as a function of barycentric distance for assumed macroscopic ring
    particle albedos of 0.04 (thick line) and 0.38 (thin line). The
    optical depth limit is derived from the normal I/F limit from
    quadrant III of the 2006 February 15 visit. The horizontal dashed
    line marks the characteristic ring optical depth predicted by
    \cite{Sternetal06}.  Orbital semi-major axes of Nix and Hydra are
    shown by vertical dotted lines at 48,675 and 64,780 km,
    respectively.
 \label{tau_vs_radius}} 
\end{figure}

Assuming that macroscopic particles dominate the backscatter from the
ring, the normal optical depth of the ring can be derived via the
following equation:

\begin{equation}
  \label{tau_eqn}
  (I/F)_{\perp}  = p \tau_{\perp}  
\end{equation}

\noindent where $p$ is the geometric albedo of the particles. To 
provide an upper limit on the normal optical depth of macroscopic ring
particles, we assume a very dark ring particle albedo of
$p_{_V}=0.04$, similar to particles in the rings of Uranus
\citep{Karkoschka01a}. We also calculate normal optical depth limits
for an assumed macroscopic ring particle albedo as high as Charon's,
i.e., $p_{_V}$=0.38 \citep{Buieetal97}. The derived upper limits on
normal optical depth are shown versus barycentric distance in
Figure~\ref{tau_vs_radius}, and the values of these upper limits at
the minimum stable barycentric distance, Nix's orbital distance, and
Hydra's orbital distance are presented in
Table~\ref{constraints_table}.  \placefigure{tau_vs_radius}

\section{Implications}

Based on the \hst observations of Pluto in February and March 2006,
the Pluto system does not currently have rings with a normal I/F in
excess of 5.1x10$^{-7}$ or a normal optical depth greater than
1.3x10$^{-5}$. This implies that the current Plutonian ring system, if
it exists, is either tenuous, like Jupiter's ring system, or is
tightly confined to less than 1500~km in width, below the spatial
resolution of our search. However, given that there are no additional
satellites of the Pluto system with an effective diameter larger than
16~km \citep{Steffletal06b}, it is difficult to imagine how such a
narrow ring could be formed or maintained. While usefully
constraining, the derived limits on the normal optical depth of
macroscopic ring particles are still somewhat higher than the typical,
time-averaged optical depth of $\tau$=5x10$^{-6}$ predicted by
\cite{Sternetal06}.

We next comment on what our results imply regarding hazards for the
{\it New Horizons} spacecraft traversing the Pluto system. Assuming a
unimodal size distribution of ring particles, the number of particles
that would impact the spacecraft is given by:

\begin{equation}
  N_p=\tau \sec \theta \left(\sigma_{NH}/\sigma_p \right)
\end{equation}

\noindent where $\theta$ is the inclination of the spacecraft 
trajectory with the ring plane, $\sigma_{NH}$ is the cross-sectional
area of the spacecraft, and $\sigma_p$ is the cross-sectional area of
a ring particle. For a ring optical depth of $\tau$=1.3x10$^{-5}$, a
trajectory inclined 30$^{\circ}$ to the ring plane, a spacecraft
projected cross sectional area of $\sigma_{NH} \approx$10~m$^2$, and a
unimodal distribution of 1~$\mu$m diameter ring particles, we find
2x10$^8$ impacts on the spacecraft, with a total impactor mass of
1x10$^{-4}$~g, assuming a ring particle density, $\rho_p$, of 1~g
cm$^{-3}$. {\it New Horizons} is robust enough to withstand collisions
with 1~$\mu$m particles at its Pluto flyby speed of $\sim$12~km
s$^{-1}$. However, collisions with 100~$\mu$m diameter particles are
potentially damaging; for a unimodal ring of 100~$\mu$m particles, up
to 2x10$^4$ such collisions could be expected.

Our ring constraints are not tight enough (by several orders of
magnitude) to ensure the safety of the {\it New Horizons} spacecraft,
were it to cross the ring plane at a distance where ring particles
might be present. In the absence of tighter constraints on the optical
depth of a putative ring system, we therefore recommend that {\it New
  Horizons} cross the ring plane within 42,000~km of the system
barycenter, where dynamical interactions with Charon prevent stable
orbits \citep{Sternetal94,Nagyetal06}. The current planned trajectory
for {\it New Horizons} is safely within this distance.

Finally, we can use our ring optical depth constraint to estimate the
lifetime of ring particles at Pluto. First, we point out that the
re-collision timescale for ring particles orbiting between Nix and
Hydra, i.e., their timescale to be reabsorbed by their parent
satellite can be computed following \cite{Burnsetal84}, yielding a
timescale of order a few 10$^3$ years. Effectively, this is an upper
limit for ring particle lifetimes since other loss processes would
only speed up the loss timescale.

Making the assumption of a steady state ring population, where
particle production and loss are in balance over long time scales, we
can derive a separate ring particle lifetime, $T_p$, based on our
derived ring particle optical depth:

\begin{equation}
  T_p=M_R / \left(dM_R / dt \right) 
\end{equation}

\noindent where $M_R$ is the ring mass, which can be computed from 
the optical depth constraint $\tau$ as:

\begin{equation}
  M_R = (8/3) \pi r_p \rho_p \tau R \, dR
\end{equation}

\noindent where $r_p$ is the characteristic ring particle radius, 
and $R$ and $dR$ are the ring's radius and width. $dM_R/dt$ can be
written very simply as:

\begin{equation}
  dM_R/dt=2 \gamma M_{sat}/T_{SS}
\end{equation}

\noindent where $M_{sat}$ is the average mass of Nix and Hydra,
$T_{SS}$ is the age of the solar system, and $\gamma$ is the
fractional mass of the satellites that have been lost to erosion.
Thus, we find:

\begin{equation}
  T_p = r_p \rho_p \tau T_{SS} R dR / \gamma r_{sat}^3 \rho_{sat}
\end{equation}

\noindent For a characteristic radius of 50~km and density of 2~g 
cm$^{-3}$ for Nix and Hydra, $\gamma$=10$^{-4}$ (consistent with Durda
\& Stern [2000]), and a unimodal ring of 1~$\mu$m diameter particles
with density 1~g cm$^{-3}$ extending from Nix's orbit to Hydra's
orbit, i.e., from 48,675--64,780~km \citep{Buieetal06}, we find a
crude ring particle lifetime constraint of $T_p$=900~yr.

\section{Conclusions}

We have used existing \hst ACS observations of the Pluto system to
derive the first constraints on the normal I/F and optical depth of
rings having a radial extent greater than 1500~km. We find a 3$\sigma$
upper limit of $I/F_{_\perp}$=5.1x10$^{-7}$ and, assuming a
macroscopic ring particle albedo of 0.04 (0.38), a 3$\sigma$ upper
limit $\tau_{_\perp}$=1.3x10$^{-5}$ (1.4x10$^{-6}$). Although higher
ring optical depths are possible for tightly confined rings less than
1500~km wide, this is unlikely given the apparent lack of additional
small satellites to shepherd the rings. The present-day optical depth
limits on the rings of Pluto are usefully constraining, but they
remain a factor of three larger than the characteristic ring optical
depth of 5x10$^{-6}$ predicted by \cite{Sternetal06}. We note that the
optical depth constraint derived from the \hst observations is not
sufficient to ensure safe passage of the {\it New Horizons} spacecraft
through the ring plane at distances of 42,000--65,000~km. Finally, in
the case of a ring in steady state, we estimate a ring particle
lifetime of 900 years.

\acknowledgments

Financial support for this work was provided by NASA through grant
numbers \mbox{GO-10427} and \mbox{GO-10774} from the Space Telescope
Science Institute, which is operated by the Association of
Universities for Research in Astronomy, Inc., under NASA contract
\mbox{NAS5-26555}. Additional support was provided by the {\it New
  Horizons Pluto-Kuiper Belt} mission. We thank M. Bullock, L. Young
and an anonymous referee for reading and commenting on this
manuscript.


\clearpage

\begin{deluxetable}{lcccc}
  \tablewidth{0pt}
  \tabletypesize{\footnotesize}
  \tablecaption{Selected Observational Parameters \label{obs_table}}
  \tablehead{\colhead{Observation Date (UT)\tablenotemark{a}} & \colhead{r (AU)} & 
    \colhead{$\Delta$ (AU)} & \colhead{$\alpha$ (deg)} & \colhead{B (deg)}}
  \startdata
  2006 Feb 15.659 & 31.07 & 31.54 & 1.59 & 37.2\\
  2006 Mar 02.747 & 31.08 & 31.31 & 1.77 & 37.5\\
  \enddata
  \tablenotetext{a}{Midpoint of observation}
  \tablecomments{$r$ is Pluto's heliocentric distance, $\Delta$ is
    Pluto's geocentric distance, $\alpha$ is Pluto's phase angle
    (Sun-Pluto-Earth angle), and $B$ is the ring plane opening angle.}
\end{deluxetable}

\begin{deluxetable}{lccc}
  \tablewidth{0pt}
  \tabletypesize{\footnotesize}
  \tablecaption{Selected 3$\sigma$ Upper Limits on the Putative Rings \label{constraints_table}}
  \tablehead{\colhead{Barycentric Distance (km)} & \colhead{(I/F)$_{\perp}$} & 
    \colhead{$\tau_{\perp}$, p=0.04} & \colhead{$\tau_{\perp}$, p=0.38}}
  \startdata
  42,000 (Min. stable dist.) & 5.1x10$^{-7}$ & 1.3x10$^{-5}$ & 1.4x10$^{-6}$ \\
  48,675 (Nix)   & 4.4x10$^{-7}$ & 1.1x10$^{-5}$ & 1.2x10$^{-6}$ \\
  64,780 (Hydra) & 2.5x10$^{-7}$ & 6.4x10$^{-6}$ & 6.7x10$^{-7}$ \\
  \enddata
\end{deluxetable}

\end{document}